\documentclass{sf2a-conf2015}
\usepackage{graphicx}
\usepackage[colorlinks, linkcolor=blue, citecolor=blue, breaklinks=true]{hyperref}
\usepackage[]{natbib}  
\usepackage{epstopdf}

\def\BibTeX{{\rm B\kern-.05em{\sc i\kern-.025em b}\kern-.08em
    T\kern-.1667em\lower.7ex\hbox{E}\kern-.125emX}}
\bibpunct{(}{)}{;}{a}{}{,}  %%%%%%%%%%%%%  A\&A bibliography style
\begin{document}

\TitreGlobal{SF2A 2015}

%%-----------------------------------------------------------------
%%      the top matter
%%

\title{The infrared signatures of very small grains \\ in the Universe seen by JWST}

\runningtitle{The infrared signatures of very small grains in the Universe seen by JWST}

\author{P. Pilleri$^{1,}$}\address{Universit\'e de Toulouse; UPS-OMP; IRAP;  Toulouse, France}\address{CNRS; IRAP; 9 Av. colonel Roche, BP 44346, F-31028 Toulouse cedex 4, France}
\author{O. Bern\'e$^{1,2}$}
\author{C. Joblin$^{1,2}$}
%\author{A. Fuente$^{3}$}\address{Observatorio Astron\'omico Nacional, Apdo. 112, 28803 Alcal\'a de Henares, Madrid, Spain}

%% IF Author3 has the same affiliation than Author1:
%\author{C.\,E. Author3$^1$}

%% IF Author3 has its own affiliation:
%\author{C.\,E. Author3}\address{Dept. of Chess, University of Games, 35101 Las Vegas, Monaco} 

%% IF Author3 has two affiliations, the one of Author1 and a second one:
%\author{C.\,E. Author3$^{1,}$}\address{Dept. of Chess, University of Games, 35101 Las Vegas, Monaco} 

%% Keep this line, even if the page will be settled afterwards.
\setcounter{page}{1}

%%-----------------------------------------------------------------

\maketitle

%%-----------------------------------------------------------------
%%        The abstract
%% 
%%  Warning!  within the abstract:
%%  - do not use macros. 
%%  - do not use commands like: \cite, \citet, \citep ... etc.

\begin{abstract}
The near- and mid-IR spectrum of many astronomical objects is dominated by emission bands
due to UV-excited polycyclic aromatic hydrocarbons (PAH) and evaporating very small grains
(eVSG).  Previous studies with the ISO, {\it Spitzer} and AKARI space telescopes have shown that the spectral variations
of these features are directly related to the local physical conditions 
that induce a photo-chemical evolution of the band carriers. 
Because of the limited sensitivity and spatial resolution, these studies have focused 
mainly on galactic star-forming regions. We discuss how the advent of JWST will allow to extend 
these studies to previously unresolved sources such as near-by galaxies, and how the analysis of the infrared signatures of PAHs and eVSGs can be used to determine their physical conditions and chemical composition. 
\end{abstract}

\begin{figure}[ht!]
 \centering
 \includegraphics[angle=270, width=1\textwidth,clip]{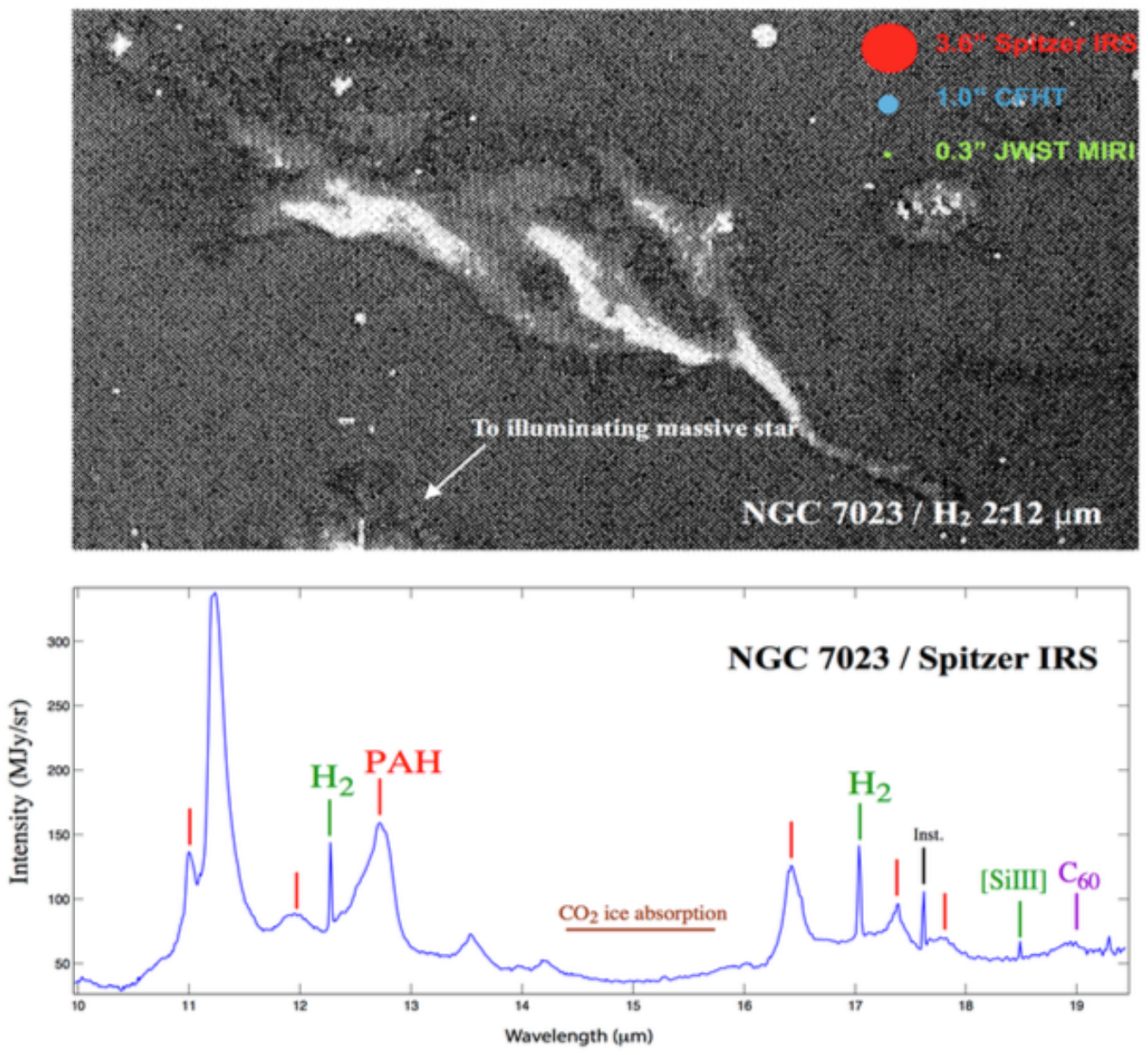}%      
%% Note the ABSENCE of the extension .pdf  !
  \caption{{\it Top:} Image of the vibrationally-excited H$_2$ line at 2.12 $\mu$m in the north-west PDR of NGC 7023 \citep{lemaire96}. The bright filamentary structures correspond to the border of the molecular cloud, illuminated by the UV photons from the massive star HD~200775. The PDR corresponds to the transition region from atomic to molecular gas, and to the zone in which very small grains are evaporated into PAHs \citep{berne07, pilleri12}. The coloured circles represent the spatial resolution of the  \emph{Spitzer}-IRS and JWST-MIRI spectrometers, and that of the H$_2$ image at 2.12\,$\mu$m. {\it Bottom:} mid-IR spectrum toward the H$_2$ filaments obtained with the \emph{Spitzer}-IRS spectrometer with average spectral resolution (R$\sim$600). The signatures of the different dust and gas populations are shown. These same signatures are often seen in different astronomical objects such as proto-planetary disks, planetary neabulae and starburst galaxies. }
    \label{fig_ngc7023}
\end{figure}

%% Insert the keywords (to appear in the ADS indexing)
%% Keywords must be separated by a comma

%%-----------------------------------------------------------------

\section{Introduction}

The infrared (IR) and (sub-)millimetre emission of star-forming galaxies is due to gas and dust in photon-dominated regions \citep[PDR, for a review, see][]{hollenbach97}. In these regions, ultraviolet (UV) photons from nearby massive stars illuminate the interstellar matter, creating a transition between the HII region  and the neutral molecular gas. 
PDRs are seen in our Galaxy  at the border of molecular clouds, in reflection nebulae, proto-planetary disks and the envelopes of evolved stars. 
To understand the shaping and evolution of these objects and, at larger scales, of starburst galaxies, it is key to understand the micro-physics that regulate the thermal and chemical balance of PDRs. Indeed, the physics and chemistry of these sources are strongly influenced by the interaction of UV photons with very small (nanometer-size) dust particles. These particles absorb most of the UV photons, re-emitting the majority of the energy in the IR domain and using a small fraction of it to heat the gas. Using the IR features to trace the evolution of the dust populations, one  can therefore hope to evaluate their impact on the thermal balance and the chemistry.

In the mid-IR domain (3-20 $\mu$m), the spectrum of PDRs presents a series of emission bands that are usually attributed to polycyclic aromatic hydrocarbons \citep[PAH; ][]{leger84, allamandola85}, evaporating very small grains \citep[eVSG; ][]{pilleri12} and to buckminsterfullerne \citep[C$_{60}$; ][]{sellgren10}, superimposed to a (generally weak) continuum. It also shows  rotational lines of molecular hydrogen \citep[H$_2$; e.g.][]{habart11}, and can also display  fine structure lines of atoms with ionisation potential lower than 13.6 eV (such as Silicon and Sulfur), as well as absorption bands  of ices at the surface of micrometer-size grains \citep{boogert15}. The ro-vibrational lines of simple molecules such as acetylene (C$_2$H$_2$) have also been detected in the dense PDRs associated with the inner envelopes of evolved stars \citep{cernicharo01} or with proto-planetary disks \citep{carr08}.  
The sub-mm and mm domains also contain a wealth of information on the morphology, dynamics and chemical richness of PDRs. In this domain, we observe the emission of   rotational lines of molecules with a permanent dipole moment, recombination lines of atomic carbon and hydrogen, and the continuum due to micrometer-size grains at thermal equilibrium. In the following, we show that the observations of both the IR and (sub-)mm domains with sufficient sensitivity and spatial resolution is an important tool in the study of galactic and extragalactic PDRs.

\section{Resolving steep chemical variations: JWST and the evaporation of very small grains}

\label{sec_chem}

Spectral mapping observations with the \emph{Spitzer} Space Telescope and with the IRAM telescopes of several PDRs have shown that the transition region between the atomic and molecular gas hosts a number of processes that determine the chemical evolution of hydrocarbon material: for instance,  IR observations have shown that eVSGs of aliphatic / aromatic composition are evaporated into free-flying PAHs \citep{berne07, pilleri12, pilleri15}. The comparison with mm observations suggests that  the photo-destruction of PAHs and eVSGs also releases small hydrocarbons in the gas phase \citep{pety05, pilleri13, guzman15, cuadrado15}. 
%The photo-evaporation process is related to fundamental mechanisms in the PDR such as the photo-electric heating efficiency \citep{bakes84, okada13} and the chemical balance. 

With the advent of ALMA, we are now peering into the details of the molecular gas in PDRs with an unprecedented angular resolution, similar to that achieved by the \emph{Hubble} Space Telescope in the visible. Figure \ref{fig_ngc7023} shows the intensity map of the vibrationally excited H$_2$ line at 2.12\,$\mu$m of the reflection nebula NGC 7023 ($d=430$\,pc) and its mid-IR spectrum obtained with the {\it Spitzer}-IRS spectrograph. This region hosts structures at different gas densities from a very diluted atomic gas at $n_H \sim 10^3$\,cm$^{-3}$ in the cavity  \citep{berne12} to dense filaments with $n_H \sim 10^5 -10^6$\,cm$^{-3}$ that are located $\sim50''$ north-west of the star. 
The filaments have angular scales of the order of 1-2$''$ \citep[e.g., ][Fig. \ref{fig_ngc7023}]{fuente96, lemaire96}. Only space-based observatories can provide enough sensitivity and spectral coverage to fully study these filaments, but until now they have been limited to an angular resolution of few arcsec. Thus, it is not yet possible to resolve the gradients in temperature, density and chemical abundances that are associated with these structures. 
%Obtaining spatially-resolved spectral imaging in the (sub-)mm and IR domains is key to understand the processes that drive the physical and chemical evolution of these objects. 
Only  JWST will be able to obtain spectral cubes of sufficiejnt sensibility, frequency coverage and spatial resolution to obtain a complete view of the chemical evolution of PAHs and eVSGs at these spatial scales and understand the processes that drive the physical and chemical evolution of these objects.

\section{The mid-IR spectrum of PDRs as tracers of the physical conditions}

The spectral properties (position, absolute and relative intensity, shape) of the infrared bands are linked to the nature of their carriers, to the physical conditions in the emitting regions (that determine the excitation conditions of the band carriers), and to the morphology of the source (that influences the radiative transfer). Thus, the analysis of the IR spectrum can be used to obtain insights on these properties. 

\begin{figure}[!b]
 \centering
  \includegraphics[height = 6cm]{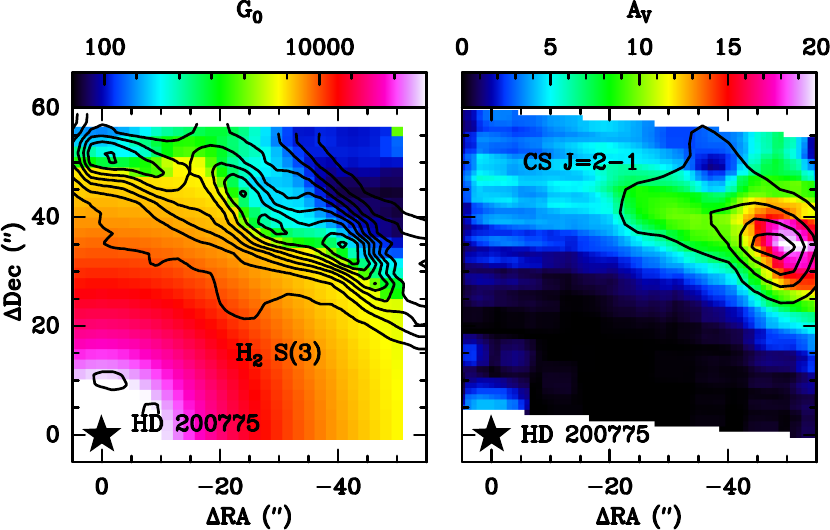}  
  \includegraphics[height =5cm]{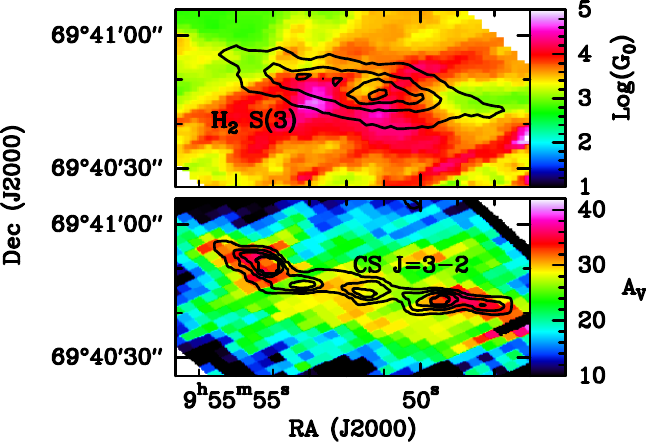}%      
%% Note the ABSENCE of the extension .pdf  !
  \caption{{\it Left:} maps of the UV radiation field intensity \citep[$G_0$; first panel, for details, see ][]{pilleri15} and of the line-of-sight extinction ($A_V$, second panel) for NGC 7023 as derived from the PAHTAT tool \citep{pilleri12} applied to the \emph{Spitzer}-IRS data cube. The contours represent the H$_2$ S(3) emission and CS $J = 2 - 1$ emission (Plateau de Bure Interferometer; Fuente et al., priv. comm.) {\it Right:} Same as the left panel, for the starburst galaxy M82. Overlaid are the H$_2$ S(3) and CS $J=3-2$ \citep{ginard15}.}
    \label{fig_pahtat}
\end{figure}

The PAHTAT toolbox \citep{pilleri14} allows to fit an observed mid-IR spectrum using a set of template spectra for ionised and neutral PAHs and eVSGs. 
In \citet{pilleri12}, we have shown that the fraction of carbon atoms contained in eVSGs ($f_C$) decreases significantly with increasing intensity of the UV radiation field, G$_0$. 
 Thus, given a mid-IR spectrum presenting emission of PAHs and eVSGs,  PAHTAT allows to estimate the intensity of the local UV field.
 We applied PAHTAT  to all the spectra contained in the {\it Spitzer}-IRS 3D cube of  NGC 7023 NW.  Figure \ref{fig_pahtat}  shows a map of G$_0$  obtained with this method \citep[for details, see][]{pilleri15}. The UV field intensity decreases by over 2 order of magnitude in only few arcsec due to the absorption by very small dust particles in the thin filaments shown in Fig. \ref{fig_ngc7023}. 
 
PAHTAT also allows to derive the selective extinction ($A_V$) by taking into  account the absorption of micron-sized grains (silicates) along the line of sight. Figure \ref{fig_pahtat} compares the map of A$_V$ with  the intensity of the  CS $J=2-1$ emission observed at the Plateau de Bure Interferometer (PdBI, A. Fuente, priv. comm.). The CS intensity presents a peak that corresponds spatially to the peak in the $A_V$ derived by PAHTAT, indicating that its emission arises from a clump of high (column) density.

We applied the same method to the {\it Spitzer}-IRS spectral cube of the starburst galaxy M~82 and obtained maps of both G$_0$ and $A_V$ (Fig. \ref{fig_pahtat}, right panel). The results show a very high extinction ($A_V > 30$) in the lines of sight close to the plane of the galaxy, with individual peaks in the NE, SW and in the center.  These values are consistent with previous analysis of mid-IR spectra obtained by \citet{beirao08}. As in the case of NGC 7023, the extinction peak correlates very well with high-density tracers such as CS $J=3-2$ \citep{ginard15}.  Because the extinction in the galactic plane is so extreme, it is difficult to probe precisely  the intensity of the UV field in the disk. However, the map of $G_0$ obtained with PAHTAT shows a clear asymmetry between the northern and southern halos, the latter presenting a higher G$_0$ compared to the former. 
This is consistent with direct observations of the far-UV luminosity of this source \citep{hoopes05} and is likely explained by a higher star formation rate in the south.

\section{Conclusions}

The analysis of the mid-IR emission of very small dust particles provides complementary information to characterise the properties of PDRs compared to molecular data in the mm for instance. We have shown that mid-IR observations can be used as a probe of physical conditions such as the UV radiation field intensity and the extinction along the observed line of sight. 

The emission of PAHs is also found to correlate with that of the fine structure line of ionised carbon  (C$^+$) at 158\,$\mu$m  \citep{joblin10}.  This line is often used as a tracer of star formation but its observation at high spatial resolution is challenging. If the PAH bands can be used as a complementary proxy to measure the star formation rate, JWST will be able to do so at an unprecedented physical scale for near-by and more distant galaxies.

%Because the IR emission of PAHs is due to the  cooling cascade after absorption of UV photons, it is often used as a tracer of star formation. The  fine structure line of ionised carbon (C$^+$) at 158\,$\mu$m, which instead traces the gas cooling, is difficult to observe at high spatial resolution. Both PAHs and C$^+$ trace the more exposed layers of the PDRs, in which only molecules that are particularly resistant to photo-dissociation (like PAHs) can survive. With the advent of JWST, the observations of PAH features will be available at an unprecedented angular resolution, allowing to peer into the star formation process of more distant galaxies. 

\begin{acknowledgements}
We thank A. Fuente for providing the CS $J = 2-1$ map of NGC 7023.\\
P. Pilleri gratefully acknowledges financial support from the Centre National d'Etude Spatiales (CNES). \\
The research leading to these results has also received funding from the European Research Council under the European Union's Seventh Framework Programme (FP/2007-2013)  ERC-2013-SyG, Grant Agreement n. 610256 NANOCOSMOS. \\
This work was supported by the CNRS program ÒPhysique et Chimie du Milieu InterstellaireÓ (PCMI) \\
The IRS was a collaborative venture between Cornell University and Ball Aerospace Corporation funded by NASA through the Jet Propulsion Laboratory and Ames Research Centre.
\end{acknowledgements}

\bibliographystyle{aa}  % A\&A bibliography style file (aa.bst)
\bibliography{pilleri} % your references in file: Yourfile.bib

\end{document}